\documentclass[twocolumn,twoside,slac,floatfix]{revtex4}
\usepackage{graphicx}
\usepackage{fancyhdr}
\begin{document}

\title{Electro and gamma nuclear physics in {\sc Geant4}}

%

\author{J.P. Wellisch}
\affiliation{CERN, Geneva, Switzerland}
\author{M. Kossov}
\affiliation{ITEP, Moscow, Russia}
\author{P. Degtyarenko}
\affiliation{TJNAF, Newport News, Virginia, USA}

\begin{abstract}
Adequate description of electro and gamma nuclear physics is of utmost importance
in studies of electron beam-dumps and intense electron beam accelerators. I also
is mandatory to describe neutron backgrounds and activation in linear colliders.
This physics was elaborated in Geant4 over the last year, and now entered into
the stage of practical application.

In the {\sc Geant4} Photo-nuclear data base there are at present 
about 50 nuclei for which the
Photo-nuclear absorption cross sections have been measured. 
Of these, data on 14 nuclei are used to parametrize the gamma nuclear reaction
cross-section  The resulting
cross section is a complex, factorized function of $A$ and 
$e = log(E_\gamma)$, where $E_\gamma$ is 
the energy of the incident photon. 

Electro-nuclear reactions are so closely connected with Photo-nuclear
reactions that sometimes they are often called ``Photo-nuclear''. The
one-photon exchange mechanism dominates in Electro-nuclear reactions, and
the electron can be substituted by a flux of 
photons. Folding this flux with the gamma-nuclear cross-section, we arrive at
an acceptable description of the electro-nuclear physics.

Final states in gamma and electro nuclear physics are described using chiral
invariant phase-space decay at low gamma or equivalent photon energies, and quark
gluon string model at high energies.

We will present the modeling of this physics in {\sc Geant4}, and show results from
practical applications.

\end{abstract}

\maketitle

\thispagestyle{fancy}


\section{Gamma nuclear reactions}
\subsection{Cross-section calculation}
The Photo-nuclear cross sections parameterized in the 
{\sc Geant4} covers all incident photon energies from
hadron production threshold up-wards.  The parameterization is subdivided 
into five energy regions, each corresponding to the physical process that 
dominates.
These are the Giant Dipole Resonance (GDR) region, the ``quasi-deuteron'' 
region, the $\Delta$ isobar region characterized by the dominant peak in the 
      cross section which extends from the pion threshold to 450 MeV,
the Roper resonance region that extends from roughly 450 MeV to 1.2 GeV, and
the Reggeon-Pomeron region above 1.2 GeV. 

From the {\sc Geant4}\cite{geant4} Photo-nuclear data base currently 14 nuclei are used in  defining
the parameterization: $^1$H, $^2$H, $^4$He, $^6$Li, $^7$Li, $^9$Be, 
$^{12}$C, $^{16}$O, $^{27}$Al, $^{40}$Ca, Cu, Sn, Pb, and U.  The result 
 is a function of $A$ and $e = log(E_\gamma)$, where $E_\gamma$ 
is the incident photon energy. The cross-section is the sum of the 
components which parameterize each energy region. \\

The cross section in the GDR region can be described as the sum of two 
peaks,
\begin{eqnarray*}
GDR(e) = th(e,b_1,s_1)\cdot exp(c_1-p_1\cdot e) + \\
         th(e,b_2,s_2)\cdot exp(c_2-p_2\cdot e) .
\end{eqnarray*}
The exponential describes the falling edge of the resonance which 
has power law behavior. The function
\begin{equation}
th(e,b_i,s_i) = \frac{1}{1+exp(\frac{b_i-e}{s_i})} ,
\end{equation}
describes the rising edge of the resonance.  It is the
nuclear-barrier-reflection function and behaves like a threshold, cutting off
the exponential.  The exponential powers $p_1$ and $p_2$ are
\begin{eqnarray*}
 p_1 = 1, p_2 = 2 \mbox{\hspace*{1mm} for \hspace*{7mm} $A < 4$ }\\ 
 p_1 = 2, p_2 = 4 \mbox{\hspace*{1mm} for \hspace*{1mm} $4 \le A < 8$ }\\
 p_1 = 3, p_2 = 6 \mbox{\hspace*{1mm} for $8 \le A < 12$} \\
 p_1 = 4, p_2 = 8 \mbox{\hspace*{1mm} for \hspace*{6mm} $A \ge 12$} .
\end{eqnarray*}
The $A$-dependent parameters $b_i$, $c_i$ and $s_i$ were found for each of
the 14 nuclei and are interpolated for other nuclei. \\

The $\Delta$ isobar region was parameterized as
\begin{equation}
\Delta (e,d,f,g,r,q)=\frac{d\cdot th(e,f,g)}{1+r\cdot (e-q)^2},
\label{Isobar}
\end{equation}
where $d$ is an overall normalization factor.  $q$ can be interpreted as the 
energy of the $\Delta$ isobar and $r$ as the inverse $\Delta$ width.  
$th$ is the threshold function.  The
$A$-dependence of these parameters is as follows:
\begin{itemize}
\item  $d=0.41\cdot A$ (for $^1$H it is 0.55, for $^2$H it is 0.88),
i.e. the $\Delta$ yield is proportional
to $A$; 
\item  $f=5.13-.00075\cdot A$.  $exp(f)$ shows how the pion threshold depends
on $A$. 
\item  $g = 0.09$ for $A \ge 7$ and 0.04 for $A < 7$;
\item  $q=5.84-\frac{.09}{1+.003\cdot A^2}$, which means that the ``mass'' 
of the $\Delta$ isobar moves to lower energies;
\item $r=11.9 - 1.24\cdot log(A)$.  $r$ is 18.0 for $^1$H.
\end{itemize}
The $A$-dependence of the $f$, $q$ and $r$ parameters is due to the 
$\Delta+N\rightarrow N+N$ reaction, which can take place in the nuclear 
medium below the pion threshold. \\

The quasi-deuteron contribution was parameterized with the same form as the 
$\Delta$ contribution but without the threshold function:
\begin{equation}
QD(e,v,w,u)=\frac {v}{1+w\cdot (e-u)^2}.
\label{QuasiD}
\end{equation}
For $^1$H and $^2$H the quasi-deuteron contribution is almost zero.  For 
these nuclei the third baryonic resonance was used instead, so the 
parameters for these two nuclei are quite different, but trivial.
The parameter values are given below.
\begin{itemize}
\item  $v = \frac {exp(-1.7+a\cdot 0.84)}{1+exp(7\cdot (2.38-a))}$, where 
$a=log(A)$.  The $A$-dependence in the quasi-deuteron 
region is stronger than $A^{0.84}$ It contributes 
only little for light nuclei. 
For $^1$H it is 0.078 and for $^2$H it is 0.08.
\item  $u = 3.7$ and $w = 0.4$.  Experimental information is 
insufficient to fix their $A$-dependence.  For both 
$^1$H and $^2$H we have $u = 6.93$ and $w = 90$.
\end{itemize}
The Roper contribution was parameterized using the same form
as the quasi-deuteron contribution:
\begin{equation}
Tr(e,v,w,u)=\frac {v}{1+w\cdot (e-u)^2}.
\label{Transition}
\end{equation}
Using $a=log(A)$, the values of the parameters are 
\begin{itemize}
\item  $v = exp(-2.+a\cdot 0.84)$.  For $^1$H it is 0.22 and for $^2$H 
it is 0.34.
\item $u = 6.46+a\cdot 0.061$ (for $^1$H and for $^2$H it is 6.57), so the 
 Roper mass increases with $A$.
\item $w = 0.1+a\cdot 1.65$.  For $^1$H it is 20.0 and for $^2$H it is 15.0.
\end{itemize}
The Regge-Pomeron contribution was parametrized in terms of two exponentials
describing the Pomeron and higher Reggeon contributions respectively:
\begin{equation}
RP(e,h)=h\cdot th(7.,0.2)\cdot (0.0116\cdot e^{(e\cdot 0.16)}
        +0.4\cdot e^{(-e\cdot 0.2)}),
\label{Regge}
\end{equation}
with $h=A\cdot exp(-a\cdot (0.885+0.0048\cdot a))$.
\subsection{Final state generation, low energies}

For simulating final states for gamma-nuclear reactions, we are using the 
chiral invariant phase-space (CHIPS) approach\cite{CHIPS1}\cite{CHIPS2}\cite{CHIPS3}.

The CHIPS model uses a set of simple rules which govern 
microscopic quark-level behavior to model macroscopic hadronic systems.  
The invariant phase space distribution
as a paradigm of thermalized chaos is applied to quarks, and simple 
kinematic mechanisms are used to model the hadronization of quarks into 
hadrons.  Along with relativistic kinematics and the conservation of quantum 
numbers, the following concepts are introduced:

The Quasmon is any excited hadronic 
system, and can be viewed as a generalized hadron.  At the constituent level, a Quasmon may be thought of as a bubble of quark-parton plasma in which the 
quarks are massless. The quark-partons in the Quasmon are massless and homogeneously 
distributed over the invariant phase space. It may also be considered as a 
bubble of the three-dimensional Feynman-Wilson \cite{Feynman-Wilson} parton 
gas.  The traditional hadron is a particle defined by quantum numbers and a 
fixed mass or width.  The quark content of the hadron is a secondary concept
constrained by the quantum numbers.  The Quasmon, however, is defined by its
quark content and mass, and the concept of a well defined particle with 
quantum numbers is of secondary importance.

The quark fusion hypothesis determines the rules 
of final state hadron production, with energy spectra reflecting the 
momentum distribution of the quarks in the system.  Fusion occurs when a 
quark-partons in a Quasmon join to form a hadron.  
In cases of multiple Quasmon, quark-partons may be exchanged between the two Quasmons.  Resulting hadrons are constrained to be 
produced on mass shell.  The type 
of the outgoing hadron is selected using combinatoric and kinematic factors 
consistent with conservation laws.

\begin{figure}
\includegraphics[angle=0,width=80mm]{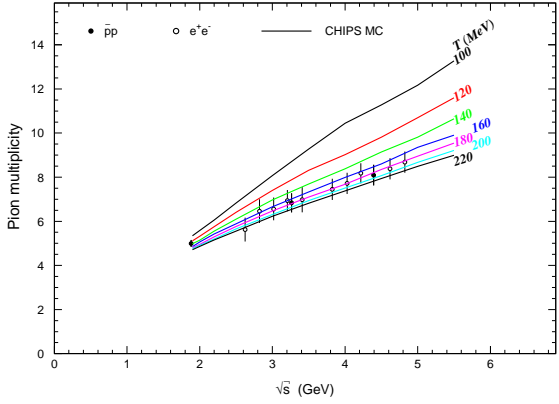}
\caption{Average meson multiplicities in proton-antiproton and in
electron-positron annihilation, as a function of the CMS energy of the
interacting hadronic system.  The points are experimental data
\protect\cite {Energy_Dep} and the lines are CHIPS Monte Carlo calculations
at several values of the critical temperature parameter $T$. }
\label{apcmul}
\end{figure}

The only non-kinematic concept 
is the hypothesis of a critical temperature of the Quasmon.  This has a 
35-year history, starting with Ref.~\cite{Hagedorn} and is based on the 
experimental observation of regularities in the inclusive spectra of hadrons
produced in different reactions at high energies.  Qualitatively, the  critical temperature hypothesis assumes that the Quasmon cannot be heated above a certain temperature.  Adding more energy to the system increases only the number of constituent 
quark-partons while the temperature remains constant.  The critical 
temperature $T_c = 180-200$ MeV is the principal parameter of the model.
The choice of this parameter is motivated from the results shown in Fig.\ref{apcmul}.

For the sake of briefness, we will only include the solution of the vacuum problem
in this paper, and refer for the solution of the in-medium equations to the CHIPS 
publications\cite{CHIPS2},\cite{CHIPS3}.
\subsubsection{Solution of the vacuum equations.} 
To generate hadron spectra from free Quasmons, the number of partons in 
the system must be found. For a finite system of $N$ partons with a total invariant mass $M$, the invariant phase space integral, $\Phi_N$, is proportional to 
$M^{2N-4}$.  At a temperature $T$ the statistical density of states is 
proportional to $e^{-\frac{M}{T}}$ and the probability to find a system 
of $N$ quark-partons in a state with mass $M$ is $dW \propto
M^{2N-4}e^{-\frac{M}{T}}dM$.  For this kind of probability distribution the 
mean value of $M^{2}$ is
\begin{equation}
<M^{2}>=4N(N-1)\cdot T^{2}.  \label{temperature}
\end{equation}
For large $N$ we obtain for massless particles the
well-known $<M>\equiv \sqrt{<M^{2}>}=2NT$ result. 

After a nucleon absorbs an incident quark-parton, such as a real or 
virtual photon, for example, the newly formed Quasmon has a total of $N$ 
quark-partons, where $N$ is determined by Eq. \ref{temperature}.
Choosing one of these quark-partons with energy $k$ in the center of mass 
system (CMS) of $N$ partons, the spectrum of the remaining $N-1$ 
quark-partons is given by
\begin{equation}
\frac{dW}{kdk} \propto (M_{N-1})^{2N-6},
\end{equation}
where $M_{N-1}$ is the effective mass of the $N-1$ quark-partons. 
The effective 
mass is a function of the total mass $M$,
\begin{equation}
M_{N-1}^{2}=M^{2}-2kM ,  \label{m_n-1}
\end{equation}
so that the resulting equation for the quark-parton 
spectrum is:
\begin{equation}
\frac{dW}{kdk}\propto (1-\frac{2k}{M})^{N-3}.  \label{spectrum_1}
\end{equation}
In order to 
decompose a Quasmon into a hadron and a residual Quasmon, one
needs the probability of two quark-partons to 
produce the effective mass of the hadron.  We calculate 
the spectrum of the second quark-parton by
following the same argument used to determine Eq.~\ref{spectrum_1}.
One quark-parton is chosen from the residual $N-1$.  It has an energy 
$q$ in the CMS of the $N-1$ quark-partons.  The spectrum is obtained by 
substituting $N-1$ for $N$ and $M_{N-1}$ for $M$ in 
Eq.~\ref{spectrum_1} and then using Eq.~\ref{m_n-1} to get 
\begin{equation}
\frac{dW}{q dq }\propto \left( 1-\frac{2q }{M\sqrt{1-
\frac{2k}{M}}}\right) ^{N-4}.  \label{spectrum_2}
\end{equation}
To ensure that the fusion will result in a  hadron of
mass $\mu$, we apply the mass shell constraint for 
the outgoing hadron,
\begin{equation}
\mu^2 = 2 \frac{k}{\sqrt{1-\frac{2k}{M}}} 
 \cdot q \cdot (1-\cos \theta ) .  \label{hadron}
\end{equation}
Here $\theta$ is the angle between the momenta {\bf k} and {\bf q} of
the two quark-partons in the CMS of $N-1$ quarks. The kinematic quark 
fusion probability for any primary quark-parton with 
energy $k$ is then:
\begin{eqnarray}
P(k,M,\mu )=&&\int \left( 1-\frac{2q }{M\sqrt{1-\frac{2k}{M}}}\right)
^{N-4} \nonumber\\
 && \times\  \delta \left( \mu ^{2}-\frac{2kq (1-\cos \theta )}{\sqrt{1-
\frac{2k}{M}}}\right) q dq d\cos \theta .\ \ \ \ 
\end{eqnarray}
Using the $\delta$-function
to perform the integration over $q$ one gets: 
\begin{eqnarray}
P(k,M,\mu )=&&\int \left( 1-\frac{\mu ^{2}}{Mk(1-\cos \theta )}\right)
^{N-4} \nonumber\\
 && \times\ \left( \frac{\mu ^{2}\sqrt{1-\frac{2k}{M}}}{2k(1-\cos \theta )}
\right)^{2}d\left(\frac{1-\cos \theta }{\mu ^{2}}\right) ,\ \ 
\end{eqnarray}
or 
\begin{eqnarray}
P(k,M,\mu )=&&\frac{M-2k}{4k}\int \left(1-\frac{\mu ^{2}}{Mk(1 - 
\cos\theta)}\right) ^{N-4} \nonumber\\
 && \times\ d\left(1-\frac{\mu ^{2}}{Mk(1-\cos \theta )}\right).
\end{eqnarray}
After the substitution
$z=1-\frac{2q }{M_{N-1}}=1-\frac{\mu ^{2}}{Mk(1-\cos \theta )}$, this
becomes
\begin{equation}
P(k,M,\mu ) = \frac{M-2k}{4k} \int z^{N-4} dz ,
\end{equation}
where the limits of integration are $0$ when 
$\cos\theta = 1 - \frac{\mu ^{2}}{M\cdot k}$, and 
\begin{equation}
z_{\max }=1-\frac{\mu^2}{2Mk}, \label{z_max}
\end{equation}
when $\cos \theta =-1$.  The resulting range of $\theta$\ is therefore
$-1<\cos \theta < 1-\frac{\mu ^{2}}{M\cdot k}$.  Integrating from $0$ to 
$z$ yields
\begin{equation}
\frac{M-2k}{4k\cdot (N-3)}\cdot z^{N-3},  \label{z_probab}
\end{equation}
and integrating from $0$ to $z_{max}$ yields the total kinematic 
probability for hadronization of a quark-parton with energy $k$ into a 
hadron with mass $\mu$:
\begin{equation}
\frac{M-2k}{4k \cdot (N-3)} \cdot z_{\max}^{N-3} . 
                                   \label{tot_kin_probab}
\end{equation}
The ratio of expressions \ref{z_probab} and \ref{tot_kin_probab} can be 
treated as a random number, $R$, uniformly distributed on the interval
[0,1].  Solving for $z$ then gives
\begin{equation}
z=\sqrt[N-3]{R}\cdot z_{\max }.  \label{z_random}
\end{equation}
In addition to the kinematic selection of the two quark-partons in the
fusion process, the quark content of the Quasmon and the spin of the 
candidate final hadron are used to determine the probability that a 
given type of hadron is produced.  Because only the relative hadron 
formation probabilities are necessary, overall normalization factors can
be dropped.  Hence the relative probability can 
be written as
\begin{equation}
P_h(k,M,\mu )=(2s_h+1)\cdot z_{\max }^{N-3}\cdot C_{Q}^{h} .  
                                         \label{rel_prob}
\end{equation}
Here, only the factor $z_{\max }^{N-3}$ is used since the other factors 
in equation \ref{tot_kin_probab} are constant for all candidates for the 
outgoing hadron.  The factor $2s_h+1$ counts the spin states of a 
candidate hadron of spin $s_h$, and $C_{Q}^{h}$ is the number of ways the 
candidate hadron can be formed from combinations of the quarks within the 
Quasmon.  In making these combinations, the standard quark wave functions 
for pions and kaons were used.  For $\eta$ and $\eta^{\prime }$ mesons the 
quark wave functions 
$\eta=\frac{\bar{u}u+\bar{d}d}{2}-\frac{\bar{s}s}{\sqrt{2}}$ and
$\eta^{\prime }=\frac{\bar{u}u+\bar{d}d}{2}+\frac{\bar{s}s}{\sqrt{2}}$
were used.  No mixing was assumed for the $\omega $\ and $\phi $\ meson 
states, hence $\omega =\frac{ \bar{u}u+\bar{d}d}{\sqrt{2}}$ and 
$\varphi=\bar{s}s$.
\subsection{Final state generation, high energies}
At high energies we use quark-gluon string model and a diffractive Ansatz for string excitation to describe the interactions of real and virtual photons with nuclei. A description of
the means of doing this can be found in a separate paper in the present proceedings\cite{Stringstuff}.
\section{Electro-nuclear scattering}
Electro-nuclear reactions are very connected with Photo-nuclear reactions. 
They are sometimes called ``Photo-nuclear'' because the one-photon 
exchange mechanism dominates the reaction.  In this sense 
electrons can be replaced by a flux of equivalent photons.  This is not 
completely true, because at high energies 
diffractive mechanisms are possible, but these types of reactions are beyond 
the scope of this discussion.
\subsection{Electro-nuclear cross-sections}
The Equivalent Photon Approximation (EPA) was proposed by
E. Fermi \cite{Fermi} and developed by C. Weizsacker and E. Williams
\cite{WeiWi} and by L. Landau and E. Lifshitz \cite{LanLif}. The
covariant form of the EPA method was developed in Refs. \cite{Pomer} and 
\cite{Grib}.  When using this method it is necessary to take into account 
that real photons are always transversely polarized while virtual photons 
may be longitudinally polarized.  In general the differential cross section
of the Electro-nuclear interaction can be written as
\begin{equation}
\frac{d^2\sigma}{dydQ^2}=\frac{\alpha}{\pi Q^2}(S_{TL}\cdot(\sigma_T
+\sigma_L)-S_L\cdot\sigma_L),
\label{elNuc}
\end{equation}
where
\begin{eqnarray*}
S_{TL}=y\frac{1-y+\frac{y^2}{2}+\frac{Q^2}{4E^2}
-\frac{m^2_e}{Q^2}(y^2+\frac{Q^2}{E^2})}{y^2+\frac{Q^2}{E^2}},\\
S_L=\frac{y}{2}(1-\frac{2m_e^2}{Q^2}).
\label{SL}
\end{eqnarray*}
The differential cross section of the Electro-nuclear scattering can be
rewritten as
\begin{equation}
\frac{d^2\sigma_{eA}}{dydQ^2}=\frac{\alpha y}{\pi Q^2}\left(\frac{(1-\frac{y}{2})^2}
{y^2+\frac{Q^2}{E^2}}+\frac{1}{4}-\frac{m^2_e}{Q^2}\right)\sigma_{\gamma^*A},
\label{difBase}
\end{equation}
where $\sigma_{\gamma^*A}=\sigma_{\gamma A}(\nu)$ for small $Q^2$ and
is written as a function of $\epsilon$, $\nu$, and $Q^2$ for
large $Q^2$.  Interactions of longitudinal photons are normally included in the 
effective $\sigma_{\gamma^*A}$ cross section through the $\epsilon$ factor, 
but in the present method, the cross section of virtual photons is 
considered to be $\epsilon$-independent.  The Electro-nuclear problem, with
respect to the interaction of virtual photons with nuclei, can thus be split 
in two.  At small $Q^2$ it is possible to use the $\sigma_\gamma(\nu)$ cross 
section.  In the $Q^2>>m^2_e$ region it is necessary to calculate the effective
$\sigma_{\gamma^*}(\epsilon,\nu,Q^2)$ cross section. \\

Following the EPA notation, the differential cross section of Electro-nuclear 
scattering can be related to the number of equivalent photons 
$dn=\frac{d\sigma}{\sigma_{\gamma^*}}$.  For $y<<1$ and $Q^2<4m^2_e$ the 
canonical method \cite{encs.eqPhotons} leads to 
\begin{equation}
\frac{ydn(y)}{dy}=-\frac{2\alpha}{\pi}ln(y).
\label{neq}
\end{equation}
In \cite{Budnev}, integration over $Q^2$ for $\nu^2>>Q^2_{max}\simeq m^2_e$
leads to
\begin{equation}
\frac{ydn(y)}{dy}=-\frac{\alpha}{\pi}\left(
\frac{1+(1-y)^2}{2}ln(\frac{y^2}{1-y})+(1-y)\right).
\label{lowQ2EP}
\end{equation}
In the $y<<1$ limit this formula converges to Eq.(\ref{neq}).  But the
correspondence with Eq.(\ref{neq}) can be made more explicit if the exact 
integral
\begin{equation}
\frac{ydn(y)}{dy}=\frac{\alpha}{\pi}\left(
\frac{1+(1-y)^2}{2}l_1-(1-y)l_2-\frac{(2-y)^2}{4}l_3\right),
\label{diff}
\end{equation}
where $l_1=ln\left(\frac{Q^2_{max}}{Q^2_{min}}\right)$,
$l_2=1-\frac{Q^2_{max}}{Q^2_{min}}$, 
$l_3=ln\left(\frac{y^2+Q^2_{max}/E^2}{y^2+Q^2_{min}/E^2}\right)$,
$Q^2_{min}=\frac{m_e^2y^2}{1-y}$, 
is calculated for $Q^2_{max(m_e)}=\frac{4m^2_e}{1-y}$.

The factor $(1-y)$ is used arbitrarily to keep $Q^2_{max(m_e)}>Q^2_{min}$, 
which can be considered as a boundary between the low and high $Q^2$ 
regions.  The transverse photon flux can be calculated as an integral 
of Eq.(\ref{diff}) with the maximum possible upper limit
\begin{equation}
Q^2_{max(max)}=4E^2(1-y).
\label{Q2max}
\end{equation}
It can be approximated by
\begin{equation}
\frac{ydn(y)}{dy}=-\frac{2\alpha}{\pi}\left(
\frac{(2-y)^2+y^2}{2}ln(\gamma)-1\right),
\label{neqHQ}
\end{equation}
where $\gamma=\frac{E}{m_e}$.  It must be pointed out that neither this
approximation nor Eq.(\ref{diff}) works at $y\simeq 1$.  
The formal limit of the 
method is $y<1-\frac{1}{2\gamma}$. \\
\begin{figure}[tbp]
   \includegraphics[angle=0,width=85mm]{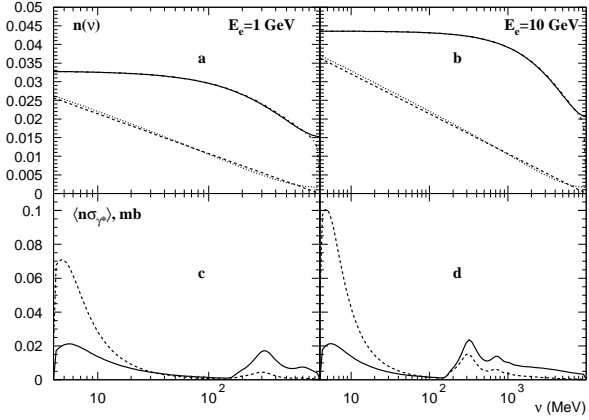}
\caption{Relative contribution of small $Q^2$
to the ``photon flux'' for (a) $1~GeV$ electrons and (b) $10~GeV$
electrons.  In figures (c) and (d) the photon distribution 
$dn(\nu,Q^2)$ is multiplied by the Photo-nuclear cross section
and integrated over $Q^2$.
The dashed lines are integrals over low $Q^2$ (under the dashed line in the first two figures), the
solid lines are integrals over high-$Q^2$ (above
the dashed lines in the first two figures).}
\label{nSigma}
\end{figure}

Fig.~\ref{nSigma}(a,b) shows the energy distribution for the equivalent photons.  The
low-$Q^2$ flux is calculated using Eq.(\ref{neq}) (dashed lines) and  Eq.(\ref{diff}) (dotted lines).  The total
flux is calculated using Eq.(\ref{neqHQ}) (the solid lines) and using
Eq.(\ref{diff}) with the upper limit defined by Eq.(\ref{Q2max}) (dash-dotted 
lines visible only around $\nu\approx E_e$).  We find that to calculate the number of low-$Q^2$ 
equivalent photons or the total number of equivalent photons one can use 
the  approximations given by Eq.(\ref{neq}) and Eq.(\ref{neqHQ}), 
respectively, instead of using Eq.(\ref{diff}). Comparing the low-$Q^2$ photon flux and the total 
photon flux we find that the low-$Q^2$ photon flux is about 
half of the the total.  From the interaction point of view the decrease of 
$\sigma_{\gamma*}$ with increasing $Q^2$ must be considered.  The 
cross section reduction for the virtual photons with large $Q^2$ is governed 
by two factors.  The cross section drops with $Q^2$ as the squared 
dipole nucleon form-factor
\begin{equation}
G^2_D(Q^2)\approx\left( 1+\frac{Q^2}{(843~MeV)^2}\right)^{-2},
\label{G2}
\end{equation}
and the thresholds the $\gamma A$ reactions are shifted to higher 
$\nu$ by a factor $\frac{Q^2}{2M}$, which is the difference between the $K$ 
and $\nu$ values.  Following the method proposed in \cite{Brasse},
$\sigma_{\gamma^*}$ at large $Q^2$ can be approximated as
\begin{equation}
\sigma_{\gamma*}=(1-x)\sigma_\gamma(K)G^2_D(Q^2)e^{b(\epsilon,K)\cdot
r+c(\epsilon,K)\cdot r^3},
\label{abc}
\end{equation}
where $r=\frac{1}{2}ln(\frac{Q^2+\nu^2}{K^2})$.  The $\epsilon$-dependence of 
the $a(\epsilon,K)$ and $b(\epsilon,K)$ functions is weak, so for simplicity 
the $b(K)$ and $c(K)$ functions are averaged over $\epsilon$.  They can be 
approximated as
\begin{equation}
b(K)\approx\left(\frac{K}{185~MeV}\right)^{0.85},
c(K)\approx-\left(\frac{K}{1390~MeV}\right)^{3}.
\label{ck}
\end{equation}
The integrated photon flux folded with the
cross section approximated by Eq.(\ref{abc}) is shown in
Fig.~\ref{nSigma}(c,d).  We show separately the low-$Q^2$ region ($Q^2<Q^2_{max(m_e)}$, dashed
lines), and the high-$Q^2$ region ($Q^2>Q^2_{max(m_e)}$, solid
lines).  These functions will be integrated over $ln(\nu)$, hence
 because of the Giant Dipole Resonance contribution, the 
low-$Q^2$ part covers more than half the total $eA\rightarrow hadrons$ 
cross section.  But at $\nu>200~MeV$, where the hadron multiplicity
increases, the large $Q^2$ part dominates.  In this sense, for a better 
simulation of the production of hadrons by electrons, it is necessary to 
simulate the high-$Q^2$ part as well as the low-$Q^2$ part. \\

\begin{figure}[tbp]
   \includegraphics[angle=0,width=85mm]{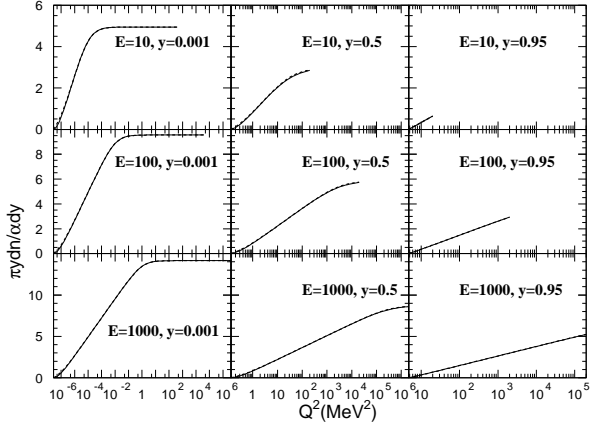}
\caption{Integrals of $Q^2$ spectra of virtual photons for three
energies $10~MeV$, $100~MeV$, and $1~GeV$ at $y=0.001$, $y=0.5$, and $y=0.95$. 
The solid line corresponds to Eq.(\protect\ref{diff}) and the dashed
line (which almost everywhere coincides with the solid line)
corresponds to Eq.(\protect\ref{diff}).}
\label{Q2dep}
\end{figure}

Taking into account the contribution of high-$Q^2$ photons it is possible to 
use Eq.(\ref{neqHQ}) with the over-estimated
$\sigma_{\gamma^*A}=\sigma_{\gamma A}(\nu)$ cross section.  The slightly 
over-estimated Electro-nuclear cross section is
\begin{equation}
\sigma^*_{eA}=(2ln(\gamma)-1)\cdot J_1-\frac{ln(\gamma)}{E_e}
\left( 2J_2-\frac{J_3}{E_e} \right).
\label{eleNucHQ}
\end{equation}
where
\begin{equation}
J_1(E_e)=\frac{\alpha}{\pi}\int^{E_e}\sigma_{\gamma A}(\nu)dln(\nu)
\label{J1}
\end{equation}
\begin{equation}
J_2(E_e)=\frac{\alpha}{\pi}\int^{E_e}\nu\sigma_{\gamma A}(\nu)dln(\nu){\rm ,\ and}
\label{J2}
\end{equation}
\begin{equation}
J_3(E_e)=\frac{\alpha}{\pi}\int^{E_e}\nu^2\sigma_{\gamma A}(\nu )dln(\nu).
\label{J3}
\end{equation}
The equivalent photon energy $\nu=yE$ can be obtained for a particular 
random number $R$ from the equation 
\begin{equation}
R=\frac{(2ln(\gamma)-1)J_1(\nu)-\frac{ln(\gamma)}{E_e}(2J_2(\nu)-\frac{J_3(\nu)}{E_e})}
{(2ln(\gamma)-1)J_1(E_e)-\frac{ln(\gamma)}{E_e}(2J_2(E_e)-\frac{J_3(E_e)}{E_e})}.
\label{RnuHH}
\end{equation}
Eq.(\ref{diff}) is too complicated for the randomization of $Q^2$ but
there is an easily randomized formula which approximates Eq.(\ref{diff})
above the hadronic threshold ($E>10~MeV$).  It reads
\begin{equation}
\frac{\pi}{\alpha D(y)}\int^{Q^2}_{Q^2_{min}}\frac{ydn(y,Q^2)}{dydQ^2}dQ^2=-L(y,Q^2)-U(y),
\label{RQ2HH}
\end{equation}
where
\begin{equation}
L(y,Q^2)=ln\left( F(y)+(e^{P(y)}-1+\frac{Q^2}{Q^2_{min}})^{-1} \right),
\label{RQ2L}
\end{equation}
\begin{equation}
D(y)=1-y+\frac{y^2}{2}{\rm ,\ and}
\label{RQ2D}
\end{equation}
\begin{equation}
U(y)=P(y)\cdot\left( 1-\frac{Q^2_{min}}{Q^2_{max}}\right) {\rm ,\ with}
\label{RQ2U}
\end{equation}
\begin{equation}
F(y)=\frac{(2-y)(2-2y)}{y^2}\cdot\frac{Q^2_{min}}{Q^2_{max}} {\rm ,\ and}
\label{RQ2F}
\end{equation}
\begin{equation}
P(y)=\frac{1-y}{D(y)}.
\label{RQ2P}
\end{equation}
The $Q^2$ value can then be calculated as
\begin{equation}
\frac{Q^2}{Q^2_{min}}=1-e^{P(y)}+\left(e^{R\cdot
L(y,Q^2_{max})-(1-R)\cdot U(y)}-F(y) \right)^{-1},
\label{Q2sol}
\end{equation}
where $R$ is a random number.  In Fig.~\ref{Q2dep}, Eq.(\ref{diff}) (solid 
curve) is compared to Eq.(\ref{RQ2HH}) (dashed curve).  Because the two 
curves are almost indistinguishable in the figure, this can be used as an 
illustration of the $Q^2$ spectrum of virtual photons, which is the derivative 
of these curves.  An alternative approach is to use Eq.(\ref{diff}) for the 
randomization with a three dimensional table $\frac{ydn}{dy}(Q^2,y,E_e)$.

\noindent 
After the $\nu$ and $Q^2$ values have been found, the value of
$\sigma_{\gamma^*A}(\nu,Q^2)$ is calculated using Eq.(\ref{abc}).
Note that if $R\cdot\sigma_{\gamma A}(\nu)>\sigma_{\gamma^*A}(\nu,Q^2)$, no
interaction occurs and the electron keeps going. 

\subsection{Final state generation}

Final states are generated using the single photon exchange assumption.
Sampling the equivalent photon distribution described in the previous section
allows to construct an exchange particle that in turn can be treated by the mechanisms 
used for gamma nuclear scattering. The question to be answered is
that of the absorption mechanisms of these particles in the context of parton 
exchange diagrams.

In the example of
the Photo-nuclear reaction discussed in the comparison section, namely
the description of $90^{\circ}$ proton and deuteron spectra in
$A({\gamma},X)$ reactions at $E_{\gamma} = 59-65$ MeV, the assumption
on the initial Quasmon excitation mechanism was the same. The
description of the $90^{\circ}$ data was satisfactory, but the
generated data showed very little angular dependence, as the velocity
of Quasmons produced in the initial state was small, and the
fragmentation process was almost isotropic.  Experimentally, the
angular dependence of secondary protons in photo-nuclear reactions is
quite strong even at low energies (see, for example,
Ref.~\cite{Ryckebusch}). This is a challenging experimental fact which
is difficult to explain in any model. It's enough to say that if the
angular dependence of secondary protons in the $\gamma ^{40}$Ca
interaction at 60 MeV is analyzed in terms of relativistic boost, then
the velocity of the source should reach $0.33 c$; hence the mass
of the source should be less than pion mass. The main subject of the present
publication is to show that the quark-exchange mechanism used in the
CHIPS model can not only model the clusterization of nucleons in nuclei
and hadronization of intra-nuclear excitations into nuclear fragments, but
can also model complicated mechanisms of interaction of photons and
hadrons in nuclear matter.

Quark-exchange diagrams
help to keep track of the kinematics of the quark-exchange process
To apply the mechanism to
the first interaction of a photon with a nucleus, it is necessary to
assume that the quark-exchange process takes place in nuclei
continuously, even
without any external interaction. Nucleons with high momenta do not
leave the nucleus because of the lack of excess energy. The
hypothesis of the CHIPS model is that the quark-exchange forces
between nucleons \cite{NN QEX}\ continuously create clusters in normal
nuclei. Since a low-energy photon (below the pion production threshold)
cannot be absorbed by a free nucleon, other absorption mechanisms
involving more than one nucleon have to be used.

\begin{figure}[tbp]
\includegraphics[angle=0,width=65mm]{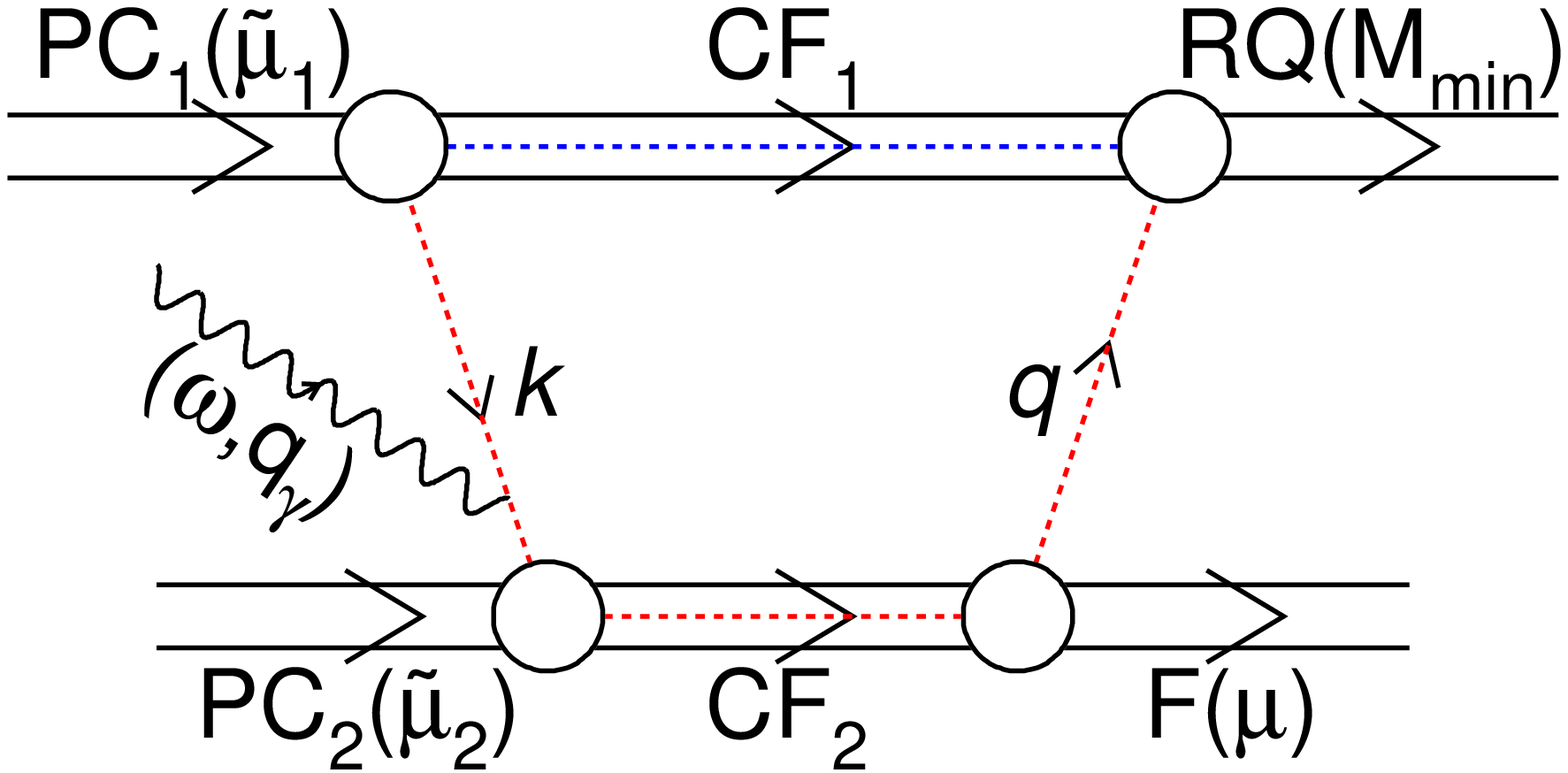}
\vskip 5mm
\includegraphics[angle=0,width=65mm]{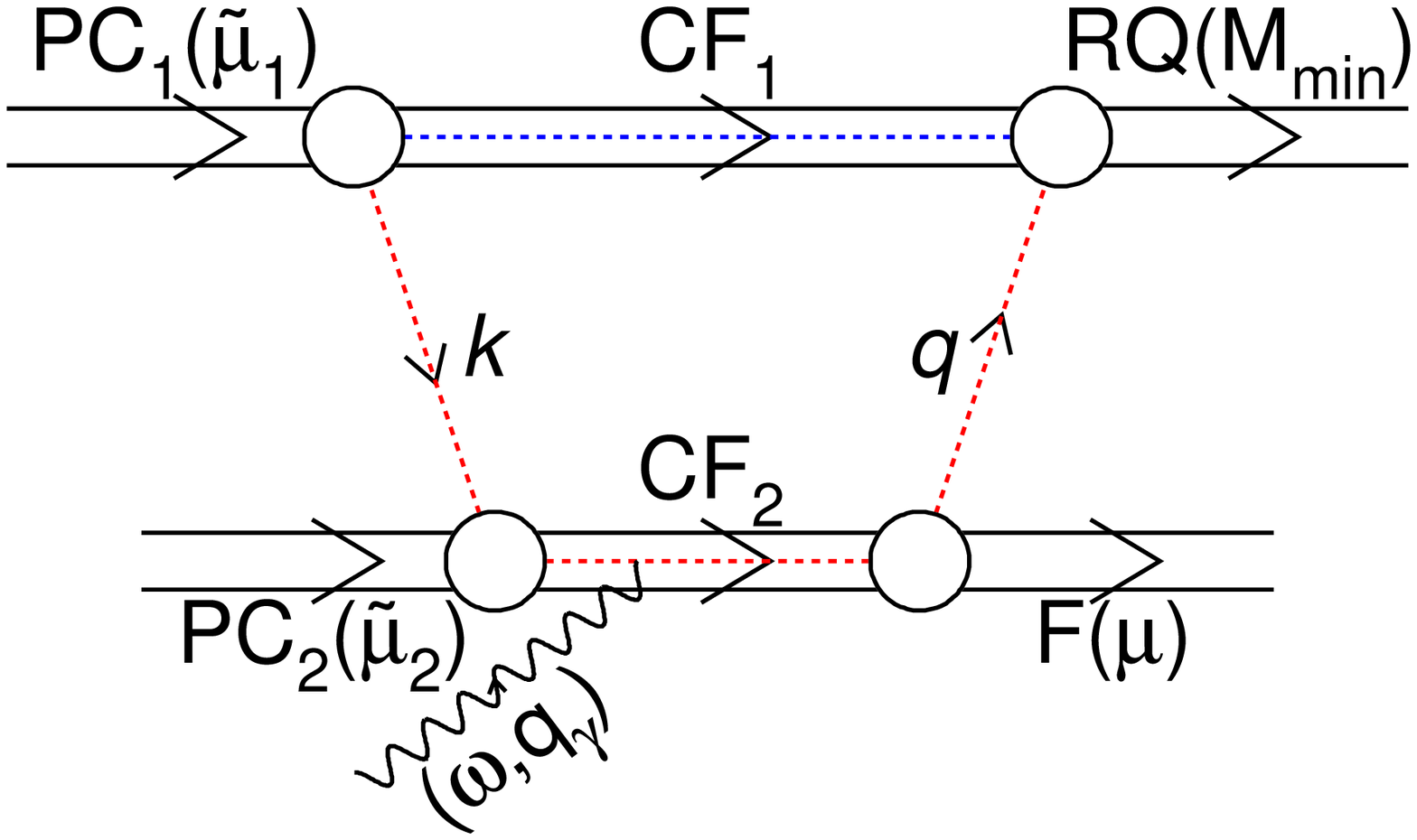}
\caption{\protect{Diagrams of photon absorption in the quark
exchange mechanism. $\rm{PC}_{1,2}$ are the parent clusters 
with bound masses
$\tilde{\mu}_{1,2}$, involved in the quark-exchange. $\rm{CF}_{1,2}$
are the colored nuclear fragments. 
F($\mu$) is the outgoing hadron with mass $\mu$. 
RQ is the residual Quasmon which carries the rest of the 
excitation energy and momentum. $M_{\min}$ characterizes 
its minimum mass defined by its quark content. Dashed lines indicate
colored objects. The photon is absorbed by a
quark-parton $k$ from the parent cluster $\rm{PC}_1$.
}}
\label{diagram1}
\end{figure}

The simplest scenario is photon absorption by a quark-parton in
the nucleon. At low energies and in vacuum this does not work because there
is no corresponding excited baryonic state. But in nuclear matter
there is a possibility to exchange this quark with a neighboring nucleon
or a nuclear cluster. The diagram for the process is shown in the upper part of
Fig.~\ref{diagram1}. In this case the photon is absorbed by a
quark-parton from the parent cluster $\rm{PC}_1$, and then 
the secondary nucleon or cluster $\rm{PC}_2$
absorbs the entire momentum of the quark and photon. The exchange
quark-parton $q$ restores the balance of color, producing the 
final-state hadron F and the residual Quasmon RQ. The process looks like a 
knockout of a quasi-free nucleon or cluster out of the nucleus. It should be
emphasized that in this scenario the CHIPS event generator
produces not only ``quasi-free'' nucleons but ``quasi-free'' fragments
too. The yield of these quasi-free nucleons or fragments is
concentrated in the forward direction.

The second scenario, shown in the lower part of Fig.\ref{diagram1} which provides for an angular dependence is the absorption
of the photon by a colored fragment ($\rm{CF}_2$ 
in Fig.~\ref{diagram1}). In this
scenario, both the primary quark-parton with momentum $k$ and the photon
with momentum $q_{\gamma}$ are absorbed by a parent cluster ($\rm{PC}_2$ in
the lower part of Fig.~\ref{diagram1}), and the recoil quark-parton with momentum $q$
cannot fully compensate the momentum $k+q_{\gamma}$. 
As a result the radiation of the
secondary fragment in the forward direction becomes more probable.

In both cases the angular dependence is defined by the first act of
hadronization. The further fragmentation of the residual Quasmon is
almost isotropic.

It was shown in above that the energy spectrum of quark
partons in a Quasmon can be calculated as
\begin{equation}
\frac{dW}{k^{\ast }dk^{\ast }}\propto 
\left(1-\frac{2k^{\ast }}{M} \right)^{N-3},  \label{spectrum_1III}
\end{equation}
where $k^{\ast }$ is the energy of the primary quark-parton in the Center
of Mass System (CMS) of the Quasmon, $M$\ is the mass of the Quasmon, 
and $N$, the number of
quark-partons in the Quasmon, can be calculated from the equation
\begin{equation}
<M^{2}>=4\cdot N\cdot (N-1)\cdot T^{2}.  \label{temperatureIII}
\end{equation}
Here $T$ is the temperature of the
system.

In the first scenario of the $\gamma A$ interaction
(Fig.~\ref{diagram1}), as both interacting particles are massless, we
assumed that the cross section for the interaction of the photon with a
particular quark-parton is proportional to the charge of the quark-parton
squared, and inversely proportional to the mass of the photon-parton
system $s$, which can be calculated as
\begin{equation}
s=2\omega k(1-\cos (\theta _{k})).  \label{s}
\end{equation}
Here $\omega $\ is the energy of the photon, and $k$ is the energy of
the quark-parton in the Laboratory System (LS):
\begin{equation}
k=k^{\ast }\cdot \frac{E_{N}+p_{N}\cdot \cos (\theta _{k})}{M_{N}}.
\end{equation}
In the case of a virtual photon, equation~(\ref{s}) can be written as
\begin{equation}
s=2k(\omega -q_{\gamma}\cdot \cos (\theta _{k})),
\end{equation}
where $q_{\gamma}$ is the momentum of the virtual photon. In both cases 
equation~(\ref{spectrum_1III}) transforms into
\begin{equation}
\frac{dW}{dk^{\ast }}\propto \left(1-\frac{2k^{\ast }}{M} \right)^{N-3},
\end{equation}
and the angular distribution in $\cos (\theta _{k})$\ converges to a
$\delta $-function: in the case of a real photon 
$\cos (\theta _{k})=1$, and
in the case of a virtual photon 
$\cos (\theta _{k})=\frac{\omega }{q_{\gamma}}$.

In the second scenario for the photon interaction 
(lower part of Fig.~\ref{diagram1}) we assumed
that both the photon and the primary quark-parton, randomized
according to 
equation~(\ref{spectrum_1III}), enter the parent cluster $\rm{PC}_2$,
and after that the normal procedure of quark exchange
continues, in which the recoiling quark-parton $q$ returns 
to the first cluster.

An additional parameter in the model is the relative contribution of
both mechanisms. As a first approximation we assumed equal
probability, but in the future, when more detailed data are obtained,
this parameter can be adjusted.

\begin{figure}[tbp]
\includegraphics[angle=0,width=65mm]{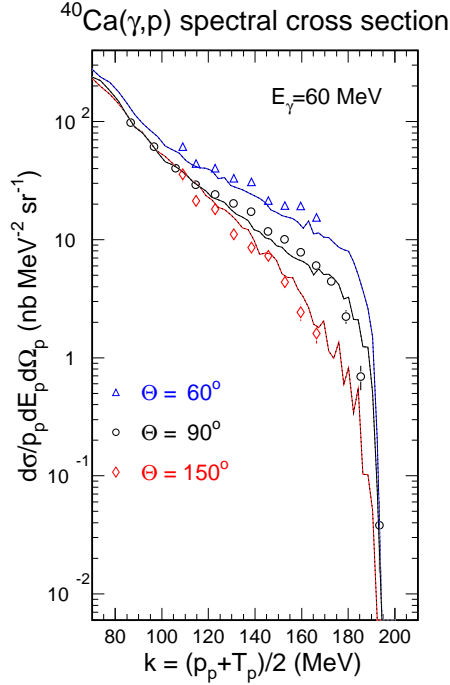}
\caption{\protect{Comparison of the CHIPS model results (lines in the
figure) with the 
experimental data~\cite{Ryckbosch} on
proton spectra at $90^{\circ}$ 
in the Photo-nuclear reactions on $^{40}$Ca at 59--65 MeV (open
circles),
and proton spectra at $60^{\circ}$ (triangles) and $150^{\circ}$ 
(diamonds).
Statistical errors in the CHIPS results are not shown but
can be judged by the point-to-point variations in the lines. The
comparison is absolute, using the value of total
Photo-nuclear cross section of 5.4 mb for Ca, 
as given in Ref.~\cite{Ahrens}.
} }
\label{gam62III}
\end{figure}

\begin{figure}[tbp]
\includegraphics[angle=0,width=65mm]{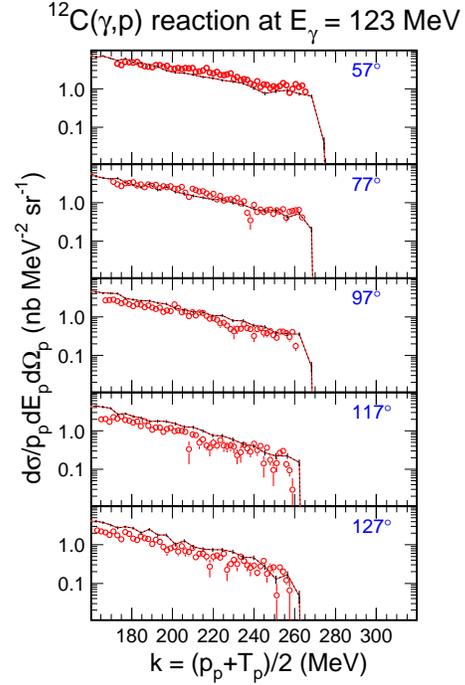}
\caption{\protect{Comparison of the CHIPS model results (lines in the
figure) with the 
experimental data~\cite{Harty} on
proton spectra at $57^{\circ}$, $77^{\circ}$, $97^{\circ}$, 
$117^{\circ}$, and $127^{\circ}$
in the Photo-nuclear reactions on $^{12}$C at 123 MeV (open
circles).
The value of the total
Photo-nuclear cross section was set at 1.8 mb.
}  }
\label{gam_123}
\end{figure}

\begin{figure}[tbp]
\includegraphics[angle=0,width=65mm]{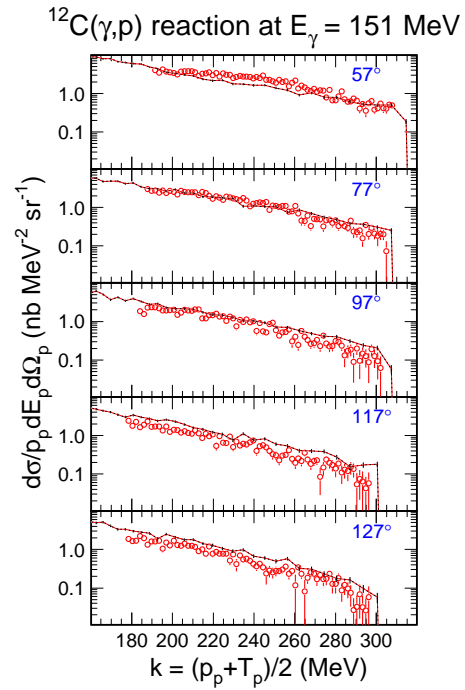}
\caption{\protect{Same as in Fig.~\ref{gam_123}, for the photon energy
151 MeV.}}
\label{gam_151}
\end{figure}

\begin{figure}[tbp]
\includegraphics[angle=0,width=65mm]{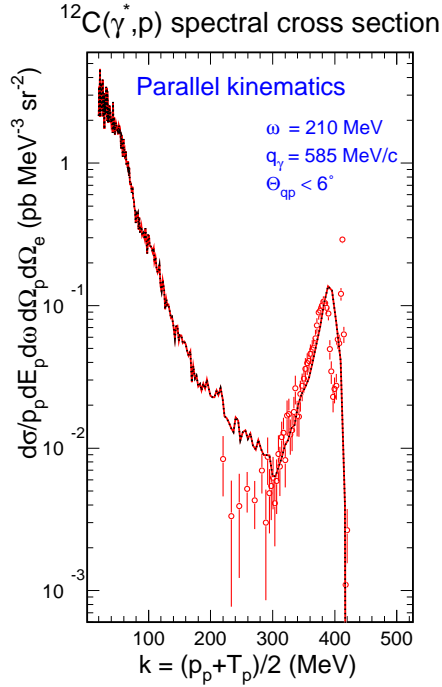}
\caption{\protect{Comparison of the CHIPS model results (line in the
figure) with the experimental data~\cite{Bates} (open circles) on the
proton spectrum measured in parallel kinematics in the
$^{12}$C(e,e$^{\prime}$p)\ reaction at an energy transfer equal to 210
MeV and momentum transfer equal to 585 MeV/$c$.  Statistical errors in
the CHIPS result are not shown but can be judged by the point-to-point
variations in the line.  The relative normalization is arbitrary.
}  }
\label{vgam}
\end{figure}
\section{Comparison with experiment}
We begin the comparison with the data on proton production in the 
$^{40}$Ca$(\gamma,X)$\ reaction at $90^{\circ}$\ at 59--65 MeV
\cite{Ryckbosch}, and at $60^{\circ}$\ and $150^{\circ}$\ at 60 MeV
\cite{Abeele}.  We analyzed these data together to compare the angular
dependence generated by CHIPS with experimental data. The data are
presented as a function of 
the invariant inclusive cross section 
$f=\frac{d\sigma }{p_{p}dE_{p}}$\ depending on the variable 
$k=\frac{T_{p}+p_{p}}{2}$, 
where $T_{p}$\ and $p_{p}$\ are the kinetic energy and the momentum of the
secondary proton. As one can see from Fig.~\ref{gam62III}, the angular
dependence of the proton yield in photo-production on $^{40}$Ca at $60$ MeV is
reproduced quite well by the CHIPS event generator.

The second set of measurements that we use for the benchmark
comparison deals with the secondary proton yields in
$^{12}$C$(\gamma,X)$ reactions at 123 and 151 MeV \cite{Harty}, 
which is still below the pion production threshold on
a free nucleon. Inclusive spectra of protons have been measured in 
$\gamma ^{12}$C reactions at $57^{\circ}$, $77^{\circ}$, $97^{\circ}$, 
$117^{\circ}$, and $127^{\circ}$. 
Originally, these data were presented as a function of 
the missing energy. We present the data in Figs.~\ref{gam_123} 
and \ref{gam_151} together with CHIPS calculations in
the form of the invariant inclusive cross section dependent on $k$. 

The agreement between the experimental data and the CHIPS model results is
quite remarkable. Both data and calculations show significant strength
in the proton yield cross section up to the kinematic limits of the
reaction. The angular distribution in the model is not as prominent as
in the experimental data, but agrees well qualitatively.

Using the same parameters, we applied the CHIPS event generator to the
$^{12}$C(e,e$^{\prime }$p) reaction measured in Ref.\cite{Bates}. The
proton spectra were measured in parallel kinematics in the interaction
of virtual photons with energy $\omega = 210$ MeV and momentum
$q_{\gamma} = 585$ MeV/$c$. To account for the experimental conditions
in the CHIPS event generator, we have selected protons generated in
the forward direction with respect to the direction of the virtual
photon, with the relative angle $\Theta_{qp} < 6^{\circ}$.  The CHIPS
generated distribution and the experimental data are shown in
Fig.~\ref{vgam} in the form of the invariant inclusive cross section as a
function of $k$.  The CHIPS event generator works only with ground
states of nuclei so we did not expect any narrow peaks for
$^{1}p_{3/2}$-shell knockout or for other shells. Nevertheless we
found that the CHIPS event generator fills in the so-called
``$^{1}s_{1/2}$-shell knockout'' region, which is usually artificially
smeared by a Lorentzian~\cite{Lorentzian}.  In the regular
fragmentation scenario the spectrum of protons below $k = 300$ MeV is
normal; it falls down to the kinematic limit. The additional yield at
$k > 300$ MeV is a reflection of the specific first act of
hadronization with the quark exchange kinematics. The slope increase
with momentum is approximated well by the model, but it is obvious
that the yield close to the kinematic limit of the $2 \rightarrow 2$
reaction can only be described in detail if the excited states of the
residual nucleus are taken into account.

The angular dependence of the proton yield in low-energy photo-nuclear
reactions is described in the CHIPS model and event generator. The
most important assumption in the description is the hypothesis of a
direct interaction of the photon with an asymptotically free quark in
the nucleus, even at low energies. This means that asymptotic freedom of
QCD and dispersion sum rules~\cite{sum_rules} can in some way be
generalized for low energies.  The knockout of a proton from a nuclear
shell or the homogeneous distributions of nuclear evaporation cannot
explain significant angular dependences at low energies.

The same mechanism appears to be capable of modeling proton yields in
such reactions as the $^{16}$C(e,e$^{\prime }$p) reaction measured at MIT
Bates \cite{Bates}, where it was shown that the region of missing
energy above 50 MeV reflects ``two-or-more-particle knockout'' (or the
``continuum'' in terms of the shell model). The CHIPS model may help
to understand and model such phenomena.

\begin{acknowledgments}
The authors wish to thank CERN for their support.
\end{acknowledgments}

\end{document}